\documentclass[prb, a4paper, twocolumn, twoside, a4paper,showpacs]{revtex4}
\usepackage{times}
\usepackage{graphicx}
\usepackage{amssymb}
\usepackage{wasysym}
\usepackage{natbib}
\usepackage[usenames,dvipsnames]{color}

\usepackage[bookmarks=true,letterpaper=true,colorlinks=true,urlcolor=blue,linkcolor=blue,citecolor=blue]{hyperref}
\begin{document}
\received{11 February 2009}
\published{22 May 2009}

\title{Influence of O$ _{\text{2}}$ and N$ _{\text{2}}$ on the
  conductivity of carbon nanotube networks} 

\author{D. J. Mowbray$^{\text{1}}$}
\email{dmowbray@fysik.dtu.dk}

\author{C. Morgan$^{\text{2}}$}

\author{K. S. Thygesen$^{\text{1}}$}

\affiliation{
$^{\text{1}}$Department of Physics, Center for Atomic-scale Materials Design (CAMD),
 Technical University of Denmark, DK-2800
Kgs.~Lyngby, Denmark\\ 
$^{\text{2}}$Department of Physics, Molecular and Materials Physics Group,  Queen Mary University of London, Mile End Road, London E1 4NS, United Kingdom
}
\begin{abstract}
We have performed experiments on single-wall carbon nanotube
  (SWNT) networks and compared with density-functional theory (DFT)
  calculations to identify the microscopic origin of the observed
  sensitivity of the network conductivity to physisorbed O$_{\text{2}}$ and
N$_{\text{2}}$. Previous DFT calculations 
of the transmission function for isolated pristine SWNTs have found
physisorbed molecules have little influence on their conductivity. 
However, by calculating the four-terminal transmission function of crossed SWNT
junctions, we show that physisorbed O$_{\text{2}}$ and N$_{\text{2}}$ do affect the junction's  conductance. This may be understood 
as an increase in tunneling probability due to hopping via molecular orbitals.
We find the effect is substantially larger for O$_{\text{2}}$ than for N$_{\text{2}}$, and for semiconducting rather than metallic
SWNTs junctions, in agreement with experiment.
\end{abstract}

\maketitle

\section{Introduction}

Using single-wall carbon nanotubes (SWNTs) as nanosensors, both
individually and in SWNT networks, has been one of the most promising
potential applications of SWNTs since their discovery.
\cite{CNTs,CNT_Sensors_Rev}  Several
experimental studies have demonstrated that the conductance of SWNT
systems is rather sensitive to the presence of even single-molecule
concentrations of physisorbed gas molecules such as O$_{\text{2}}$ and
N$_{\text{2}}$. \cite{Ref01,Ref03,Ref10, O2_Plasma,SWNT_Networks_O2,
  ChrisO2,ChrisN2}  Further, by measuring
conductivity of individually characterized SWNTs, \cite{CNT+NH3+NO2} as
well as thick (metal-like) and thin (semiconductor-like) SWNT networks,
\cite{LRef3,ChrisO2,ChrisN2} the response of SWNTs to contaminants has
been shown to
correlate with the intrinsic electronic properties of the material.
For example, it has been found that the
presence of low-O$_{\text{2}}$ concentrations, independent of
temperature, introduces an increase in conductance of approximately
20\% on thin SWNT networks, while an increase in
conductance of only about 1\% is found for thick SWNT
networks. \cite{ChrisO2}

On the other hand, previous theoretical studies have found that SWNTs
are rather inert, so that gases tend only to physisorb to the SWNT
surface.
\cite{O_Adsorption_Graphite_NTs,O2_Adsorption_SWCNTs,O_Chemisorption_CNTs,CNT+O2_10_0,CNT+O2_physi1,CNT+O2_physi2,CNT+O2_Triplet,CNT+O2_8_0} 
For this reason, it was
suggested that O$_{\text{2}}$
should not effect conductance through SWNTs, but only influence
conductance at either SWNT-SWNT junctions, at the
SWNT-metal contacts, or at SWNT defect sites.
\cite{O_Adsorption_Graphite_NTs,Sensors} Although the conductivity of
SWNTs with molecules physisorbed at defect sites has been extensively studied,
\cite{CNT_Doped_Theory,O2_Defects,CNT_Defects_Conductance}
the conductivity of four-terminal SWNT-SWNT junctions has been
previously studied only for small pristine metallic SWNTs.
\cite{ExpCrossedCNTs,Junction_Conductance} The possible influence of
physisorbed molecules on SWNT-SWNT junctions has not been investigated.

\begin{figure}[!b]
\includegraphics[width=0.9\columnwidth]{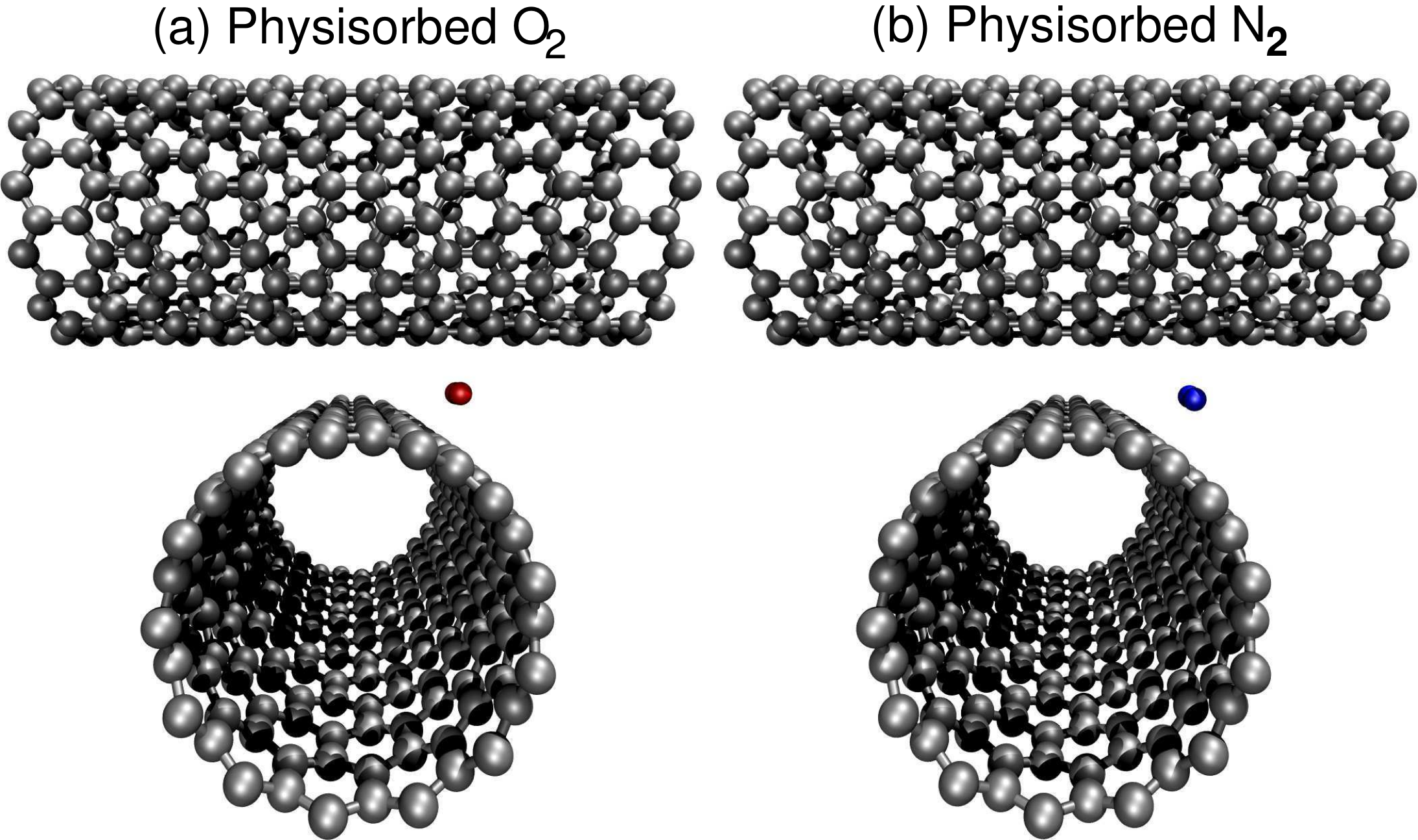}
\caption{(Color online) Schematics of a (13,0) SWNT junction with (a) physisorbed
  O$_{\text{2}}$  and (b) physisorbed N$_{\text{2}}$.
}\label{Fig1}
\end{figure}

In this paper we address the microscopic origin of the increase in conductance of SWNT networks when exposed to O$_{\text{2}}$ or N$_{\text{2}}$ gas. To this end, we have performed density-functional theory (DFT) 
calculations  of the intratube transmission
within a SWNT and the intertube transmission between two SWNTs
in the nonequilibrium Green's function (NEGF) formalism for
O$_{\text{2}}$ and N$_{\text{2}}$ molecules physisorbed in (7,7)
metallic armchair, (12,0) semimetallic zigzag, and (13,0)
semiconducting zigzag SWNT junctions, shown schematically in
Fig.~\ref{Fig1}.  Comparing our theoretical results for SWNT
  junctions with experimental measurements for SWNT networks suggests
  that the surprising sensitivity to O$_{\text{2}}$ and N$_{\text{2}}$ 
 may be partially due to an increased tunneling
probability through O$_{\text{2}}$ and N$_{\text{2}}$ physisorbed at
SWNT junctions.  

In Sec. \ref{Experimental_Results} we describe
experimental measurements of the influence of both O$_{\text{2}}$ and
N$_{\text{2}}$ on the conductivity of SWNT networks and the
characterization of these networks using Raman spectroscopy. A
description of the DFT and NEGF model used to describe the microscopic
origin of this effect is then provided in Sec. \ref{Basic_Theory}. In
Sec. \ref{Theoretical_Results_and_Discussion} we compare our
theoretical results for the SWNT junction transmission with the SWNT
network experiments, followed by a concluding section.

\section{Experimental Results}
\label{Experimental_Results}

\begin{figure}
\includegraphics[width=\columnwidth]{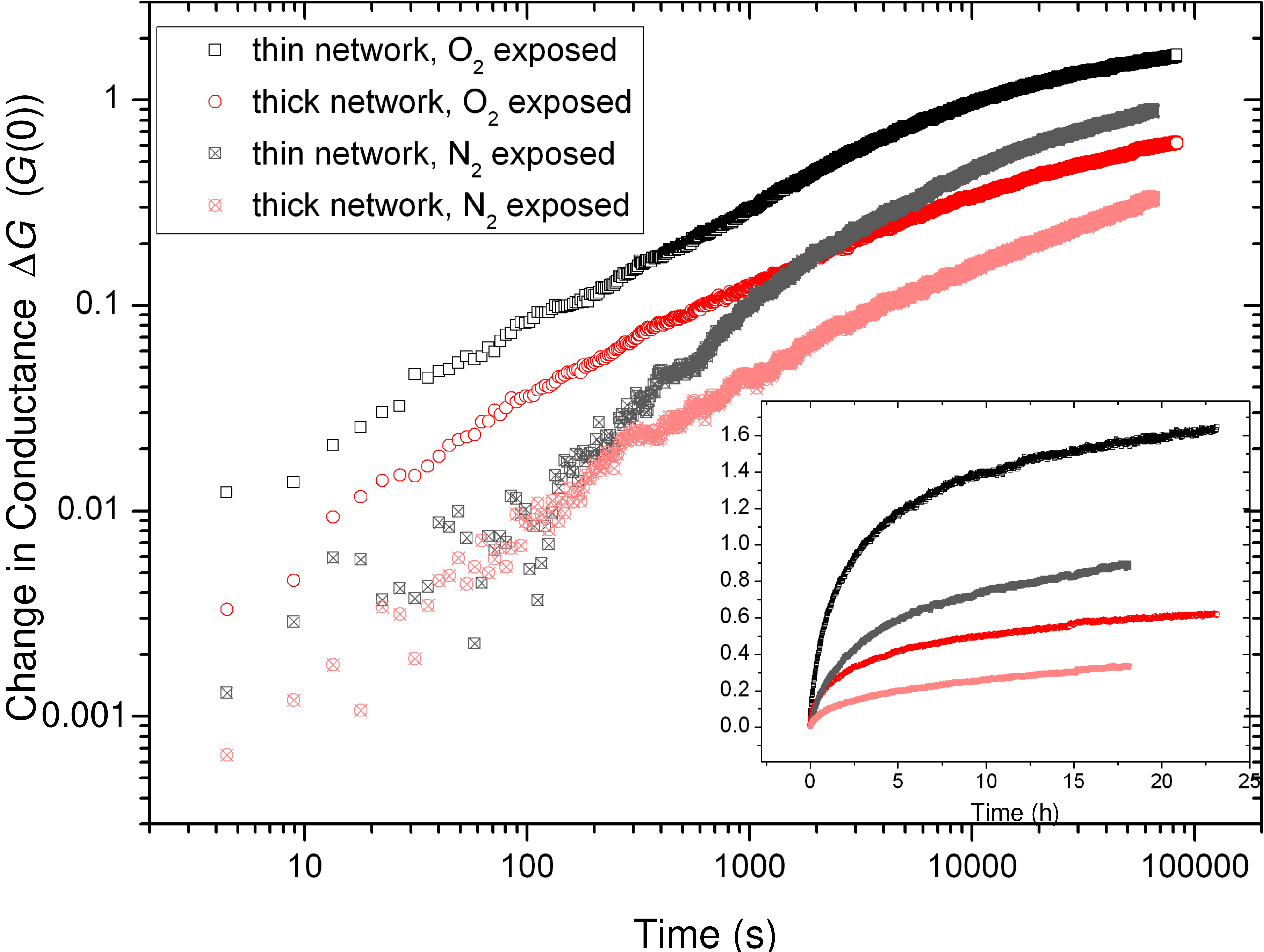}
\caption{(Color online) Fractional change in conductance $\Delta G =
    G/G(\text{0}) -\text{1}$  
  versus time $t$ in seconds and hours (inset) following exposure to
  O$_{\text{2}}$ and N$_{\text{2}}$ for thin
  ($\square,{\color{Gray}{\boxtimes}}$) and thick
  (${\color{red}{\ocircle}},{\color{Salmon}{\otimes}}$) SWNT networks
  respectively, 
  on log-log and linear (inset) scales (Refs.~\onlinecite{ChrisO2} and \onlinecite{ChrisN2}).  
}\label{Fig2}
\end{figure}

Below we give a brief discussion of our experiments on SWNT network conductivity.  A more detailed description may be found in Refs.~\onlinecite{ChrisO2} and \onlinecite{ChrisN2}.  
Figure \ref{Fig2} shows experimental measurements of the
conductance sensitivity  to O$_{\text{2}}$ and N$_{\text{2}}$
exposure for thick and thin SWNT networks.  These samples were
initially placed under vacuum ($\sim \text{1}\times
\text{10}^{-\text{6}}$ mbar) and irradiated by a UV light-emitting diode (LED) ($\lambda
\sim$ 400 nm) at low intensity ($\sim$0.03 mW/cm$^{\text{2}}$) for
approximately 12 h to desorb surface and interbundle adsorbates
(surface dopants) from the SWNTs. Once the SWNT network's
conductance stabilized, the samples were exposed to either
O$_{\text{2}}$ (99.5\% pure) or N$_{\text{2}}$ (99.998\% pure) at 1
atm.  The conductance of the samples was then monitored by
periodically sampling ($\Delta t \approx$ 1 s) the current while
applying a fixed bias of 1 mV to the thick (metal-like) SWNT network
($R \approx $ 1 k$\Omega$) and 10 mV to the thin
(semiconductor-like) SWNT network ($R \approx$ 1000 k$\Omega$), as
shown in Fig.~\ref{Fig2}. 

After 5 min of exposure to O$_{\text{2}}$, the thin
network shows an increase in conductance of about 13\% while the
thick network's conductance changes by about 7\%.  For the same
exposure to N$_{\text{2}}$, both networks show substantially smaller
conductance changes of 2\%--3\%.  However, at exposure
times of more than 2 h, the thin SWNT network response to
N$_{\text{2}}$ is similar to that of the thick SWNT network to
O$_{\text{2}}$.  This might be caused by a weaker physisorption
  of N$_{\text{2}}$ to the SWNT networks than O$_{\text{2}}$.  The inset of
Fig.~\ref{Fig2} also shows that at very long exposure times the
fractional change in conductance, \(\Delta G = G/G(\text{0}) -
\text{1}\), becomes saturated after 24 h.  Further, the response
to O$_{\text{2}}$ depicted in Fig.~\ref{Fig2} shows that the
conductance change for a thin SWNT network is about two to three
times that of the thick SWNT network at all times.  This suggests
that the conductance change under O$_{\text{2}}$ exposure is an
intrinsic property of the SWNT networks, present even at very low O$_{\text{2}}$
concentrations.  Herein we shall focus on the microscopic origin of the
network sensitivity to O$_{\text{2}}$ and N$_{\text{2}}$, with the temporal
behavior of the networks discussed elsewhere. \cite{ChrisO2,ChrisN2}

\begin{figure}
\includegraphics[width=\columnwidth]{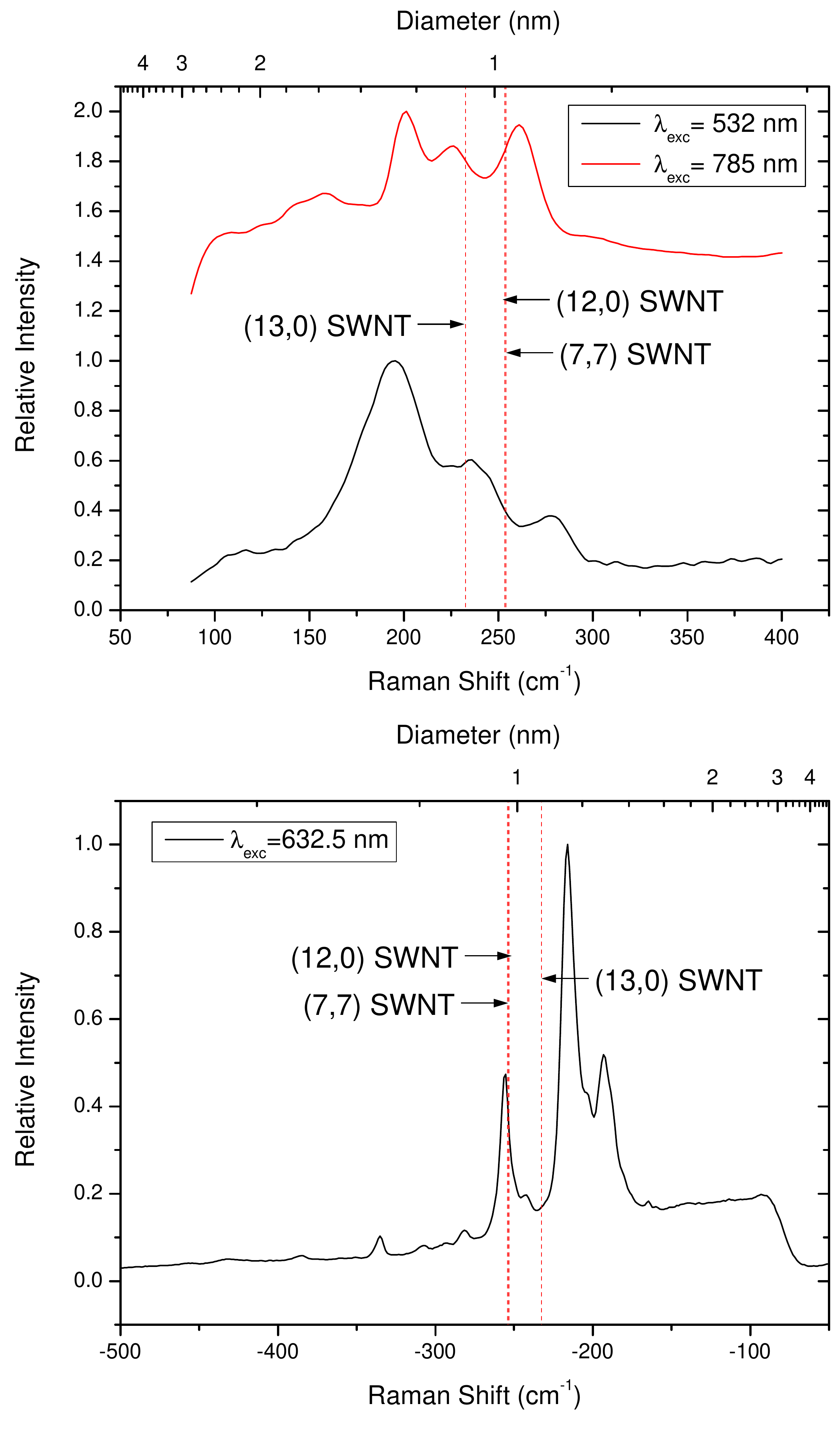}
\caption{(Color online) Raman spectra and approximate diameter distribution of HiPco
  SWNT sample for an excitation wavelength (top)
  $\lambda_{\text{exc}} \approx$ 532 nm (lower black curve),
  $\lambda_{\text{exc}} \approx$ 785 nm
  (upper red curve), and (bottom)
  $\lambda_{\text{exc}} \approx$ 632.5 nm (\textbf{---$\!$---}).  The DFT
  calculated diameters of \(d\approx\) 9.76 \AA, 9.79 \AA, and 10.66
  \AA~for (7,7), (12,0), and (13,0) SWNTs, respectively, are provided
  for comparison (dashed lines). 
}\label{Fig3}
\end{figure}

We have performed the Raman spectroscopy to characterize our SWNT
  network samples, which were produced via the high-pressure carbon
  monoxide (HiPco) method. 
Figure \ref{Fig3} shows the radial breating mode (RBM) Raman signals of HiPco samples
at excitation wavelengths  $\lambda_{\text{exc}} \approx$ 532 nm,
$\lambda_{\text{exc}} \approx$ 632.5 nm and $\lambda_{\text{exc}}
\approx$  785 nm.  The van Hove singularity energy separation was
calculated using the tight-binding approximation with the
carbon-carbon interaction energy $\gamma_0 \approx$ 2.9 eV and
carbon-carbon bond length $a_{\text{C}-\text{C}} \approx$ 1.44 \AA.
The SWNT diameter $d$ dependence of the RBM frequency
$\nu_{\text{RBM}}$ for isolated SWNTs on  SiO$_{\text{2}}$ has been shown
\cite{raman_nanotube} to behave as $\nu_{\text{RBM}} \approx
\text{248}/d_{\text{t}}$.  The DFT calculated diameters for (7,7),
(12,0), and (13,0) SWNTs of $d \approx$ 9.76, 9.79, and
10.66 \AA, respectively, are found to correlate well with the HiPco
Raman shift, as shown in Fig.~\ref{Fig3}.  This should ensure a good
description of the SWNT network's work function, which may be
significantly different for smaller tubes.

\section{Basic Theory}
\label{Basic_Theory}

\begin{figure}
\includegraphics[width=0.9\columnwidth]{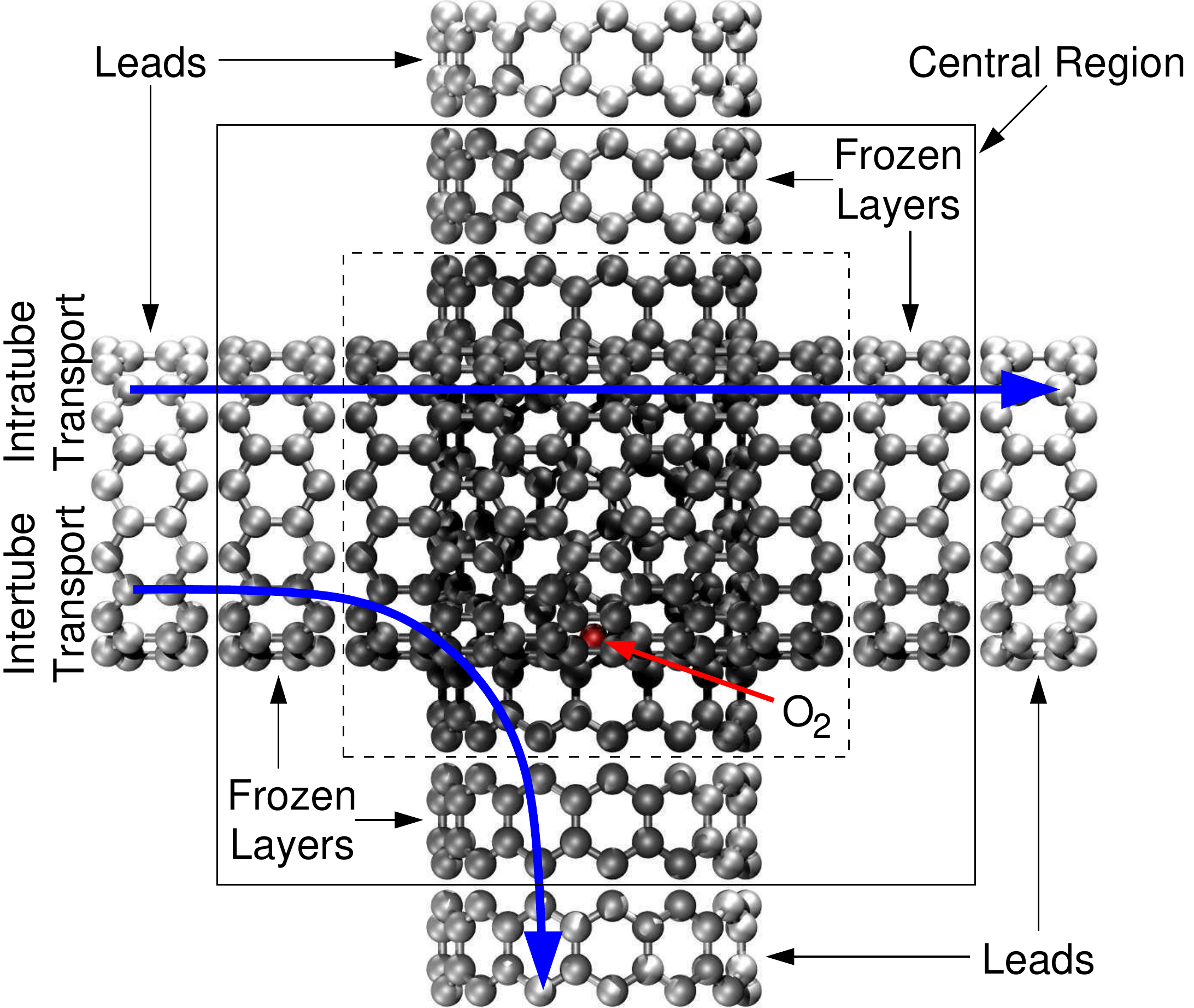}
\caption{(Color online) Schematic of (13,0) SWNT junction consisting of four SWNT
  leads (light gray) coupled to the central
  scattering region via four SWNT primitive unit cells or layers
  (gray) ``frozen'' at their relative positions in the isolated SWNT.
  The atomic positions of the remaining eight SWNT layers (dark gray)
  and the physisorbed O$_{\text{2}}$ molecule have been relaxed. A
  SWNT separation consistent with experiments of approximately 3.4
  \AA~has been used.  The intratube transport through the SWNT and the
  intertube transport between the SWNTs has been shown schematically.
}\label{Fig4}
\end{figure}

Our DFT calculations have been performed with the
{\sc{siesta}} DFT code \cite{ASE,SIESTA} using a double-zeta polarized
(DZP) basis set for the physisorbed molecules (O and N), and a single-zeta
polarized (SZP) basis set for the SWNTs (C), and the Perdew-Burke-Ernzerhof
exchange-correlation functional. \cite{PBE}  We note here that the
DZP and SZP {\sc{siesta}} basis sets have recently been shown to yield
transmission functions in quantitative agreement with plane-wave codes
and maximally localized Wannier functions. \cite{BenchmarkPaper}  When
modeling O$_{\text{2}}$ we have performed spin-polarized
calculations \cite{O_Adsorption_Graphite_NTs} but have performed spin-
unpolarized calculations otherwise.

We have modeled the junction system using 6(11) primitive unit
cells or layers for each zigzag(armchair) SWNT per supercell, with a
separation of approximately 3.4 \AA, as
depicted for a (13,0) SWNT junction in Figs.~\ref{Fig1} and \ref{Fig4}.
\cite{Jesper} The four SWNT layers at the boundaries of the central
region, shown in gray in Fig.~\ref{Fig4}, were kept fixed at their
relaxed positions in the isolated SWNT.  At the same time the central
4(9) primitive unit cells from each tube, shown in dark gray in
Fig.~\ref{Fig4}, and the physisorbed molecules were relaxed until a
maximum force of less than 0.1 eV/\AA\ was obtained.  Since the
supercell has  dimensions of  \(\gtrsim\text{25}\) \AA\ for each SWNT
junction, a \(\Gamma\) point calculation was sufficient to describe
the periodicity of the structure.

Such a large supercell was necessary for the Hamiltonian
 of each of the four SWNT layers adjacent to the
boundaries \(H_{C}^{\text{prin}}\), to be within 0.1 eV of the Hamiltonian for the
respective leads \(H_\alpha\), so that \(\max | H_{C}^{\text{prin}} -
H_\alpha| <\) 0.1 eV.  In this way the electronic structure at the
edges of the central region was ensured to be converged to that in the leads.

\begin{table}
\caption{Change in intertube conductance $\Delta
  G_{\text{inter}}$ at the valence-band maximum
  \(\varepsilon_{\text{VB}}\) relative to the pristine junction
    for O$_{\text{2}}$ and 
   N$_{\text{2}}$ physisorbed in SWNT junctions of (7,7), (12,0), and
  (13,0) SWNTs, with the respective SWNT-O$_{\text{2}}$ and
  SWNT-N$_{\text{2}}$ separations $d$ in Angstrom.}\label{Table1}
\begin{ruledtabular}
 \begin{tabular}{lcccc}
 & \multicolumn{2}{c}{$\Delta G_{\text{inter}}$ (\%)} & \multicolumn{2}{c}{$d$
    [$\text{\AA}$]}\\\cline{2-3} \cline{4-5}
SWNT Junction & \multicolumn{1}{c}{O$_{\text{2}}$} & \multicolumn{1}{c}{N$_{\text{2}}$}  &
SWNT-O$_{\text{2}}$ & SWNT-N$_{\text{2}}$ \\\hline
 (7,7) Armchair & 30 & 6 & 2.3 & 2.8 \\
 (12,0) Zigzag  & 140 & 14 & 2.6 & 2.8 \\
 (13,0) Zigzag  & 1800 & 130 & 2.5 & 2.8
 \end{tabular}
\end{ruledtabular}
\end{table}

The Landauer-B\"{u}tticker conductance for a multi-terminal system can
be calculated from the Green's function of the central region,
\(G_C\), according to the formula \cite{Meir,Thygesen2,4TConductance} 
\begin{eqnarray}
G = \left. G_0
  \mathrm{Tr}[G_C\Gamma_{\text{in}}G_C^\dagger\Gamma_{\text{out}}]\right|_{\varepsilon = \varepsilon_F}, 
\end{eqnarray}
where the trace runs over all localized basis functions in the central
region.  To describe the conductance at small bias for semiconducting
systems, the Fermi energy \(\varepsilon_F\) should be taken as the
energy of the valence-band maximum \(\varepsilon_{\text{VB}}\) or
conduction-band minimum \(\varepsilon_{\text{CB}}\) for $p$-type and
$n$-type semiconductors, respectively. The central region Green's
function is calculated from  
\begin{eqnarray}
G_C(\varepsilon) = \left[z S_C - H_C -
\sum_\alpha \Sigma_\alpha(\varepsilon)\right]^{-1},
\end{eqnarray}
where \(z = \varepsilon + i 0^+\), \(S_C\) and \(H_C\) are the overlap matrix and
Kohn-Sham Hamiltonian matrix of the central region in the localized
 basis, \(\Sigma_{\alpha}\) is the self-energy of lead $\alpha$,  
\begin{eqnarray}
\Sigma_\alpha(\varepsilon) = [z S_{C\alpha} -
  H_{C\alpha}] \left[z S_\alpha -
  H_\alpha\right]^{-1} [z S_{C\alpha}^\dagger
- H_{C\alpha}^\dagger],
\end{eqnarray}
and the coupling elements between the central region and lead $\alpha$
for the overlap and Kohn-Sham Hamiltonian are $S_{C\alpha}$ and
$H_{C\alpha}$ respectively.

The coupling strengths of the input and output leads are then given by
 \(\Gamma_{\text{in}/\text{out}} = i(\Sigma_{\text{in}/\text{out}} - \Sigma_{\text{in}/\text{out}}^\dagger)\).
For a four-terminal SWNT junction, the intratube and intertube
transmission functions are calculated by choosing the appropriate
output lead, as depicted in Fig.~\ref{Fig4}.

\section{Theoretical Results \& Discussion}
\label{Theoretical_Results_and_Discussion}

\begin{figure*}
\includegraphics[width=0.7\textwidth]{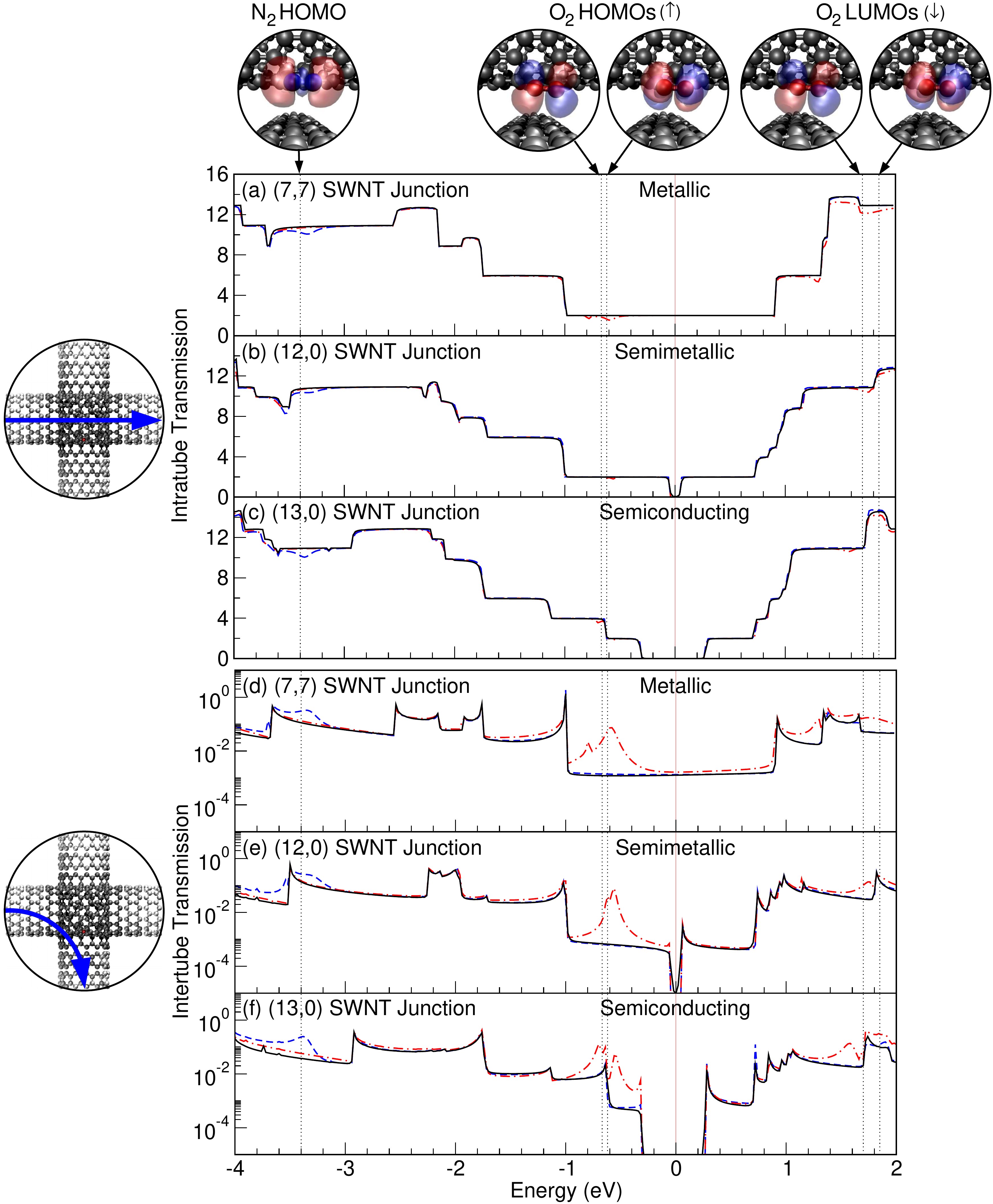}
\caption{(Color online) (a)--(c) Intratube transmission  and (d)--(f) intertube
  transmission vs energy in eV relative to the Fermi energy for a pristine
  (black solid curves) SWNT junction, and with N$_{\text{2}}$
  (blue dashed curves), and O$_{\text{2}}$
  (red dash-dotted curves) physisorbed
  consisting of [(a) and (d)] metallic (7,7) SWNTs, [(b) and (e)] semimetallic (12,0)
  SWNTs, and [(c) and (f)] semiconducting (13,0) SWNTs.  Schematics and
  isosurfaces of $\pm\text{0.02}e$/\AA$^{\text{3}}$ for molecular
  orbitals on the physisorbed molecules are shown above the
  corresponding eigenenergies (dotted lines).}\label{Fig5}
\end{figure*}

For each of the three types of SWNT junctions considered, we find that
both O$_{\text{2}}$ and N$_{\text{2}}$ are physisorbed with binding
energies of \(\sim\) 0.2 eV, as depicted in Fig.~\ref{Fig1}.
Further, the SWNT--O$_{\text{2}}$ and SWNT--N$_{\text{2}}$ equilibrium
separation distance $d$ is in the range 2.3---2.8 \AA, as given in Table
\ref{Table1}.  These results agree qualitatively with previous
theoretical studies for O$_{\text{2}}$ binding distances and
  energies on isolated SWNTs.
\cite{O_Adsorption_Graphite_NTs,O2_Adsorption_SWCNTs,O_Chemisorption_CNTs,CNT+O2_10_0,CNT+O2_physi1,CNT+O2_physi2,CNT+O2_Triplet,CNT+O2_8_0}

Figures \ref{Fig5}(a)-\ref{Fig5}(c) show the intratube transmission for three
prototypical SWNTs commonly found in experimental HiPco samples,
\cite{Photoluminescence} as shown in Fig.~\ref{Fig3}.
In Fig.\ \ref{Fig5}(a) we see that for a metallic armchair (7,7) SWNT,
transmission occurs through two channels at the Fermi level. We see in
Fig.\ \ref{Fig5}(b) that the conductance for the
semimetallic zigzag (12,0) SWNT resembles that found in
Fig.\ \ref{Fig5}(a) for a metallic SWNT, except for a tiny band gap of
\(\lesssim\) 0.05 eV at the Fermi level.  In Fig.\ \ref{Fig5}(c) we
find for a semiconducting zigzag (13,0) SWNT a band gap of
approximately 0.6 eV between the valence and conduction bands, through which
no transmission occurs. This is only slightly smaller than the expected
band gap of \(\sim\) 0.7 eV, based on a \(d^{-\text{1}}\) fit to
experimental data.\cite{ExpDOS_10_0}  These results for the
intratube transmission of pristine SWNTs also agree qualitatively
with previous DFT studies of isolated (5,5), (10,10), (10,5),
(11,0), and (12,0) SWNTs.
\cite{DOS_1010_105,Junction_Conductance,Rev_CNTTransport} We also
find in Figs.~\ref{Fig5}(a)--\ref{Fig5}(c) that neither O$_{\text{2}}$ nor
N$_{\text{2}}$ physisorbed at a SWNT junction noticeably influence the 
intratube transmission. 

In Fig.~\ref{Fig5} we also show isosurfaces and eigenenergies for the
highest occupied and lowest occupied molecular orbitals (HOMO and
LUMO) on physisorbed O$_{\text{2}}$ and N$_{\text{2}}$.  For these
weakly coupled molecules, the renormalized molecular levels may easily
be identified with the molecular orbitals of the free
O$_{\text{2}}$ and N$_{\text{2}}$ molecules. \cite{Thygesen1}
Since the position of the molecular levels is rather insensitive to
the type of junction, it should also be insensitive to the exact
binding geometry.  This suggests that additional physisorbed molecules
will influence the intertube transmission similarly. 

We find the intertube transmission is proportional to the density of
states (DOS) for the system with peaks in the transmission at the van Hove
singularities.  This is consistent with transport between the SWNTs
occurring in the tunneling regime, as expected for a SWNT separation
of approximately 3.4 \AA. 

The presence of physisorbed molecules in the SWNT-SWNT gap should then
increase the tunneling probability at energies near the eigenenergies
of the molecular orbitals.  This is evidenced by the distinct peaks in the
intertube transmission for each SWNT junction at energies corresponding
to the HOMO of N$_{\text{2}}$ and the spin polarized HOMOs and LUMOs of
O$_{\text{2}}$, as seen in Figs.~\ref{Fig5}(d)--\ref{Fig5}(f).  Under such conditions,
a SWNT junction behaves as a simple tunneling electron microscopy
(TEM) tip.  By applying appropriate bias voltages, one may potentially
probe the molecular orbitals of a physisorbed molecule to determine
its chemical composition.

For this reason, the sensitivity of SWNT network conductivity to
O$_{\text{2}}$ is at least partly due to the close proximity  of the
O$_{\text{2}}$ HOMOs to the Fermi energies of typical SWNTs
(\(\approx\) 0.6 eV), as shown in Fig.~\ref{Fig5}.  Further, it has
been shown experimentally that defects inherent in physically realizable SWNTs
yield $p$-type semiconductors.\cite{CNTs,p-type,p-type_n-type}  The
conductivity measured experimentally at small bias is thus at the
energy of the valence band \(\varepsilon_{\text{VB}}\).  

As seen in Fig.~\ref{Fig5}(f), the O$_{\text{2}}$ HOMO eigenenergy is
only about 0.3 eV below \(\varepsilon_{\text{VB}}\) for a
semiconducting (13,0) SWNT junction.  As shown in Table \ref{Table1},
this yields a substantial increase in the intertube conductance at
zero bias for semiconducting junctions in the presence of
O$_{\text{2}}$, while much smaller increases are found for the
metallic and semimetallic junctions, in agreement with experiment.
\cite{ChrisO2}  On the other hand, we also find physisorbed N$_{\text{2}}$
increases the intertube conductance only slightly, also in qualitative
agreement with experiment, \cite{Ref10} as shown in Fig.~\ref{Fig2}.

Although it is well-known DFT calculations underestimate band gaps,
\cite{DOS_1010_105,UnderestimateBG2,UnderestimateBG1} since we are primarily interested in how the
presence of O$_{\text{2}}$ or N$_{\text{2}}$ qualitatively changes
the DOS and conductance, such calculations are still useful.

\section{Conclusions}
\label{Conclusions}

In conclusion, we have proposed a possible microscopic explanation for the
experimentally observed sensitivity of the electrical conductance of
carbon nanotube networks to oxygen and nitrogen gases. Our DFT
calculations suggests that O$_{\text{2}}$ and N$_{\text{2}}$ physisorbed in crossed SWNT junctions 
can have a large influence on the intertube conductance. In particular,
for O$_{\text{2}}$ the close proximity of the highest occupied molecular orbitals with
the Fermi level of the SWNT significantly increases electron tunneling
across the gap. The effect is found to be larger for O$_{\text{2}}$ than for N$_{\text{2}}$ and
for semiconducting rather than metallic SWNTs, in agreement with the
experimental observations. Our results suggests that the electrical
properties of SWNT networks are to a large extent determined by crossed SWNT junctions.

\acknowledgments

We thank E. I. Kauppinen, and K. W. Jacobson for
useful discussions.  D.J.M. and K.S.T. acknowledge
financial support from NABIIT and the Danish Center for Scientific
Computing under Grant No. HDW-1103-06. The Center for Atomic-scale
Materials Design (CAMD) is sponsored by the Lundbeck Foundation.
C. M. acknowledges financial support from the National Physical
Laboratory (NPL) and the Engineering and Physical Sciences Research
Council (EPSRC-GB).


\end{document}